\documentclass{article}[12pt,a4paper]

\usepackage{amsmath}

\begin{document}

\title{The Expectation Value of the Cosmological Constant}

\author{Josily Cyriac\thanks{Department of Physics, Government College, Kattappana, Idukki, Kerala, India - 685508, Now at Department of Basic Sciences, Government Engineering College, Painavu, Idukki, Kerala, India - 685603}}

\date{}

\maketitle

\begin{abstract}

The possibility of extending the quantum mechanical superposition principle to free parameters such as the cosmological constant appearing in the Lagrangian of physical theories is examined. If the cosmological constant is subject to a quantum mechanical superposition principle, its observed value is a weighted average of its two natural values, one at the Planck scale and the other at zero. As zero and nonzero values of the cosmological constant leads to topologically distinct spacetimes, the amplitude for the cosmological constant to take a Planck scale value is weighted over a topological Euclidean action and the expectation value of the cosmological constant is found to be of the order of its present observed value.

\end{abstract}

\section{Introduction}

Observations during the last two decades have revealed that the universe is expanding at an accelerating rate.\cite{1999ApJ-Perlmutter}\cite{1998AJ....116.1009R}\cite{Komatsu:2010fb}\cite{Ade:2013zuv} According to the General Theory of Relativity, such an expansion is possible only if the matter content of the universe is dominated by a component with negative pressure.\cite{Peebles:2002gy} The most natural candidate for such a component of the energy-momentum tensor of the universe is a cosmological constant. However, the present value of the cosmological constant, as inferred form observations of type Ia supernovae and the cosmic microwave background radiation, is about $10^{-122}$ times too small compared to its natural value at the Planck scale.\cite{Weinberg:2000yb}  It would be natural if the cosmological constant is either exactly zero or as large as the Planck scale. However, the observed very small value of the cosmological constant is quite unnatural that requires a very large amount of fine tuning. The difficulty to explain such a tremendous fine tuning has led to the suggestion that the value of the cosmological constant is drawn to zero by some symmetry or physical mechanism and the accelerating expansion of the universe is caused by an exotic component of the energy-momentum tensor of the universe called dark energy.

Quantum mechanically, if there are two possible states, say, $\left| \alpha_1 \right\rangle$ and $\left| \alpha_2 \right\rangle$ for a system, the system can exist in a superposition state $a_1 \left| \alpha_1 \right\rangle +\  a_2 \left| \alpha_2 \right\rangle$. If $A$ is a dynamical variable of which $\left| \alpha_1 \right\rangle$ and $\left| \alpha_2 \right\rangle$ are eigenstates, the expectation value of $A$ can be defined as 
	\[
	\left< A \right> = \frac{|a_1|^2 A_1 + |a_2|^2 A_2}{|a_1|^2 + |a_2|^2}
\]
 where $A_1$ and $A_2$ are the eigenvalues of $A$ corresponding to $\left| \alpha_1 \right\rangle$ and $\left| \alpha_2 \right\rangle$ respectively. If this quantum mechanical superposition principle is extendable to free parameters such as the cosmological constant appearing in the Lagrangian of physical theories, the expectation value of such parameters can be defined as a weighted average of their naturally possible values. In this paper, the possibility for the cosmological constant to have a small expectation value is examined. 

\section{Extending the superposition principle to free parameters in the Lagrangian}

The cosmological constant can enter into the Einstein equations in two different ways, either as a constant $\lambda$ multiplying the metric tensor $g_{\mu \nu}$ or as the energy density of the vacuum. This is due to the fact that the energy-momentum tensor of the vacuum can be expressed as $T_{\mu \nu} = \rho_{vac} \ g_{\mu \nu}$ where $\rho_{vac}$ is the energy density of the vacuum. Thus a vacuum energy density of $\rho_{vac}$ is equivalent to a cosmological constant 
\begin{equation}
\lambda = 8 \pi G \ \rho_{vac}
\end{equation}
According to quantum field theory, the zero point excitations of all fields contribute to the vacuum energy and the value of the vacuum energy density is determined by the cut-off scale of the theory, which is normally taken as the Planck scale.\cite{Weinberg:1988cp}\cite{Padilla:2015aaa}\cite{1972gcpa.book.....W} Thus, the natural value of the cosmological constant is $\lambda \sim m_p^2$ where $m_p$ is the Planck mass. Even if we assume a cut of at GUT or electroweak scales, the value of the resulting cosmological constant is still many orders of magnitude higher than the observed value. As a remedy for this discrepancy, various physical mechanisms have been suggested such as the vacuum energy being canceled by some symmetry or that it does not appear in the Einstein equations.\cite{Martin:2012bt}

However, such mechanisms can exclude the vacuum energy from appearing in the Einstein equations, but not the bare cosmological constant $\lambda$. Although it is quite possible that the cosmological constant $\lambda$ is exactly zero, it is equally possible that it is of the order of unity, $\lambda \sim m_p^2$. If the cosmological constant is exactly zero, some form of dark energy such as a scalar field is required to explain the accelerated expansion of the universe. Observations have established that the dark energy component of the energy-momentum tensor of the universe is consistent with the presence of a cosmological constant.\cite{2014A&A...568A..22B} Although $\lambda = 0 $ and $\lambda \sim m_p^2$ are quite natural, the observed value $\lambda \sim 10^{-122} m_p^2$ is highly unnatural. 
 
If we consider $\lambda$ as a classical quantity, it is not subject to any natural values, as there are no natural length scales in classical physics. Quantum mechanically, there exists a natural length scale - the Planck length - and the natural value for $\lambda$ is $\sim m_p^2$. 
Usually, the values of such parameters as $\lambda$ appearing in the Lagrangian of a physical theory is not subject to the quantum mechanical superposition principle. The quantum mechanical superposition principle applies to state kets only and not to free parameters of the theory. However, as $\lambda$ is one of the very few such parameters appearing in fundamental theories, the test for such a possibility lies in comparing its outcomes with observations. 

This leads us to consider the possibility that $\lambda$ is a quantum mechanical parameter that can take two possible values, say, $\lambda_0$ and $\lambda_1$ corresponding to two different states of the metric tensor $g_{\mu \nu}$, the observed value of $\lambda$ being an average or expectation value corresponding to a superposition of the two states. 

For simplicity, we consider a universe which is devoid of any matter other than the cosmological constant. A value of $\lambda = 0$ correspond to a flat universe whereas a value of $\lambda \sim m_p^2$ correspond to a de-Sitter universe with a Hubble constant $H_0 \sim m_p$. If we work in Euclidean spacetime, $\lambda = 0$ correspond to a flat spacetime and $\lambda \sim m_p^2$ correspond to a four-sphere of radius $\sim l_p$. Thus, a transition from $\lambda=0$ to $\lambda \sim m_p^2$ involves a change in the topology of the universe. If we consider $\lambda$ as a quantum mechanical parameter, such a change is possible through a quantum tunneling process, the amplitude for which can be expressed as $\sim e^{-S}$, where $S$ is the corresponding action. If $S \gg 1$, the probability for $\lambda \sim m_p^2$ is very small, $e^{-S}$, and this would lead to a small expectation value of the order of $ \sim m_p^2 \ e^{-S}$ for the cosmological constant. 

\section{The expectation value of the cosmological constant}

Einstein equations with a non-zero cosmological constant is given by
\begin{equation}
	R_{\mu \nu} - \frac{1}{2} g_{\mu \nu} R - \lambda g_{\mu \nu} = 8 \pi G \ T_{\mu \nu}
\end{equation}
Classically, there are no restrictions on the value of $\lambda$ as there are no natural length scales in classical physics. However, quantum mechanically, there exists a natural length scale 
		\[
	l_p^2 \sim \frac{\hbar G}{c^3}
\]
and we expect $\lambda$ to be of the order of $\frac{1}{l_p^2} = m_p^2$, if not zero.

In an otherwise empty universe, $\lambda = 0$ leads to Einstein equations
\begin{equation}
	R_{\mu \nu} - \frac{1}{2} g_{\mu \nu} R = 0
\end{equation}
the solution of which is the flat spacetime
	\[
	g_{\mu \nu} = \eta_{\mu \nu}
\]
where $\eta_{\mu \nu}$ is the Minkowski metric. $\lambda \neq 0$ leads to Einstein equations
\begin{equation}
	R_{\mu \nu} - \frac{1}{2} g_{\mu \nu} R - \lambda g_{\mu \nu} = 0
\end{equation}
the solution of which is the de-Sitter spacetime, a coordinate representation of which is given by
	\[
	ds^2 =  -\left[ 1- \frac{r^2}{S^2} \right] dt^2 + \frac{dr^2}{\left[ 1- \frac{r^2}{S^2} \right]} + r^2 d\theta^2 + r^2 sin^2 \theta d\phi^2
\]
where $S^2 = \frac{3}{\lambda}$. A value of $\lambda \sim m_p^2$ leads to $S \sim l_p$.

In Euclidean spacetime, $\lambda \neq 0$ gives a flat spacetime, 
	\[
	g_{\mu \nu} = \delta_{\mu \nu}
\]
and $\lambda \neq 0$ leads to a four sphere
\begin{equation}
	ds^2 = S^2 \left[ d\theta^2 + sin^2 \theta d\phi^2 + sin^2 \theta sin^2 \phi  d\psi^2 + sin^2 \theta sin^2 \phi sin^2 \psi d\chi^2 \right]
\end{equation}
where $S^2 = \frac{3}{\lambda}$. A value of $\lambda \sim m_p^2$ again leads to $S \sim l_p$.

Classically, only one of these two possibilities can be actually realized, but, quantum mechanically there are different possibilities. 

a) The universe can initially be in a state with $\lambda = 0$ and can tunnel into a state with $\lambda \neq 0$. The amplitude for such a tunneling process is $\sim e^{-S}$ where $S$ is the action associated with the tunneling process. 

b) The universe can initially in a state with $\lambda \neq 0$ and can tunnel into a state with $\lambda = 0$. 

A third possibility arises if we assume that $\lambda$ is subject to a quantum mechanical superposition principle.

c) Instead of tunneling between two topologically district universes, the universe can exist in a state with a value of $\lambda$, which is an average of its natural values. 

To find the average or expectation value of the cosmological constant, the amplitudes for $\lambda = 0$ and $\lambda \neq 0$ need to be determined. If $S$ is the action associated with the tunneling process from $\lambda = 0$ state to $\lambda \neq 0$ state, the amplitude for $\lambda \neq 0$ state is $\sim e^{-S}$. If $S$ is sufficiently large, $e^{-S} \ll 1$ and the amplitude for the $\lambda = 0$ state is $\sim 1$.

With $\lambda \sim m_p^2$, this leads to an expectation value of 
\begin{equation} \label{eqn:eqna}
	\left< \lambda \right> = m_p^2 \ e^{-S}
\end{equation}
Thus, if $S \gg 1$, it is possible for the cosmological constant to have an exponentially small value. 

To calculate the value of $\lambda$, the action $S$ appearing in equation (\ref{eqn:eqna}) need to be determined. The obvious choice is the Einstein-Hilbert action\cite{1977PhRvD..15.2752G}\cite{hawking1978quantum}
\begin{equation}
	S = \frac{1}{16 \pi G} \int d^4 x \sqrt{g} \ (R - 2\lambda)
\end{equation}
Even in Euclidean spacetime, this action is not finite when the spacetime is not compact. To get a finite value of the cosmological constant, it required that the action be finite. As the tunneling process from a $\lambda = 0$ state to a $\lambda \neq 0$ in Euclidean spacetime involves a change in the topology of spacetime, we consider the possibility that $S$ is a topological action.

It may be noted that the solution for $g_{\mu \nu}$ in Euclidean spacetime with $\lambda \neq 0$ is a gravitational instanton, a four-dimensional complete regular manifold with a positive-definite metric.\cite{PhysRev.97.511}\cite{1978NuPhB.144..349H} It has been argued that gravitational instantons may play a role in solving the cosmological constant problem.\cite{1998CQGra..15.2629C}\cite{1993CQGra..10..207C}\cite{1988NuPhB.310..643C}\cite{1997PhRvL..79.4071C} Although the action for gravitational instantons is usually taken to be the Einstein-Hilbert action, here, we try to construct a topological action.

One reason for considering a topological action in the functional integral for the cosmological constant is the fact that the Yang-Mills instanton action given by\cite{1995qtf..book.....W}
\begin{equation}
	S = \int d^4 x \ F^a_{\mu  \nu} F_a ^{\mu  \nu}
\end{equation}
can be interpreted as a) the action of field theory and b) as a topological term related to the Pontryagin index
\begin{equation}
	\sigma = \frac{1}{8 \pi^2} \int d^4 x \ F^a_{\mu  \nu} F_a ^{\mu  \nu}
\end{equation}
In Yang-Mills theory, the action of the theory is identical with the topological term but for certain numerical constants. However, the Einstein-Hilbert action cannot be easily identified with such topological invariants. 

Thus, it may be possible that the action $S$ appearing in equation (\ref{eqn:eqna}) is not the Einstein-Hilbert action, but a topological term that can be expressed in terms of certain topological invariants. The most natural candidate for such a topological invariant for gravitational instantons is the Euler characteristic\cite{2009arXiv0912.4922P}
\begin{equation}
	\chi = \frac{1}{128 \pi^2}\int d^4 x \ \sqrt{g} \ R_{abcd} \ R_{efgh} \ \epsilon^{abef} \ \epsilon^{cdgh}
\end{equation}
Then, the action $S$ appearing in equation (\ref{eqn:eqna}) is
\begin{equation}
	S = \frac{1}{\alpha^2}\int d^4 x \ \sqrt{g} \ R_{abcd} \ R_{efgh} \ \epsilon^{abef} \ \epsilon^{cdgh}
\end{equation}
where $\alpha$ is a normalization constant. The observed value of the cosmological constant is given by 
\begin{equation} 
	\left< \lambda \right> = \lambda e^{-\frac{1}{\alpha^2}\int d^4 x \ \sqrt{g} \ R_{abcd} \ R_{efgh} \ \epsilon^{abef} \ \epsilon^{cdgh}}
\end{equation}
where the integral is over the four sphere $S_4$.

This may be further generalized to a functional integral as 
\begin{equation} 
	\left< \lambda \right> = \int D[g_{\mu \nu}] \ \lambda e^{-S}
\end{equation}
where the functional integration is over all possible compact four manifolds that satisfy the Euclidean Einstein equation with a cosmological constant.
\begin{equation}
	G_{\mu \nu} = \Lambda \ g_{\mu \nu}
\end{equation}
This integral can be evaluated using the steepest descent approach, the dominant contribution to the functional integral arises from the topology of compact four manifolds for which the integral 
\[
	\int d^4 x \ \sqrt{g} \ R_{abcd} \ R_{efgh} \ \epsilon^{abef} \ \epsilon^{cdgh}
\]
 is minimum, which is the four sphere whose Euler number is $\chi=2$.

To calculate the value of the observed cosmological constant, the value of the normalization constant $\alpha$ also need to be known. The Euler number can be expressed in terms of the Gauss-Bonnet invariant
\begin{equation}
	\chi = \frac{1}{32 \pi^2}\int d^4 x \ \sqrt{g} \ (R_{\mu \nu \rho \delta} \ R^{\mu \nu \rho \delta} - 4 R_{\mu \nu} \ R^{\mu \nu} + R^2)
\end{equation}
For a four sphere of radius $S$, a single term in the sum $R_{\mu \nu} R^{\mu \nu}$ contribute $\frac{9}{S^4}$. Therefore we chose the normalization constant as $\alpha ^2 = 9$. Thus
\begin{equation} 
	\left< \lambda \right> = \lambda e^{-\frac{1}{9}\int d^4 x \ \sqrt{g} \ R_{abcd} \ R_{efgh} \ \epsilon^{abef} \ \epsilon^{cdgh}}
\end{equation}
As the Euler characteristic for the four-sphere has a value of $2$, the integral 
\[
	\int d^4 x \ \sqrt{g} \ R_{abcd} \ R_{efgh} \ \epsilon^{abef} \ \epsilon^{cdgh}
\]
has a value of $256 \pi ^2$ and with  $\lambda \sim m_p^2$
\begin{equation} 
	\left< \lambda \right> \sim m_p^2 \ e^{-\frac{256 \pi^2}{9}} 
\end{equation}
or 
\begin{equation} 
	\left< \lambda \right> \sim 10^{-122} \ m_p^2  
\end{equation}
which is consistent with observations.\cite{2009ApJ...699..539R}\cite{2012ApJ...746...85S} 

\section{Conclusion}

It has been shown that extending the quantum mechanical superposition principle to free parameters appearing in the Lagrangian of physical theories leads of a small value for the cosmological constant. If this is true, the accelerated expansion of the universe can be interpreted as a low energy observable outcome of quantum effects in gravity.

\end{document}